# Hydrogen spillover and storage on graphene with single-site Ti catalysts

Jhih-Wei Chen[1,2,#], Shang-Hsien Hsieh[2,#], Sheng-Shong Wong[1,#], Ya-Chi Chiu[1], Hung-Wei Shiu[2], Chia-Hsin Wang[2], Yaw-Wen Yang[2], Yao-Jane Hsu[2], Domenica Convertino[3], Camilla Coletti[3], Stefan Heun[4], Chia-Hao Chen[2*], and Chung-Lin Wu[1,2*]

[1]Department of Physics, National Cheng Kung University, Tainan 70101, Taiwan

[2]National Synchrotron Radiation Research Center, Hsinchu 30076, Taiwan

[3]Center for Nanotechnology Innovation @ NEST, Istituto Italiano di Tecnologia, Pisa 56127, Italy

[4]NEST, Istituto Nanoscienze-CNR and Scuola Normale Superiore, 56127 Pisa, Italy

[#]Contributed equally

*Correspondence and requests for materials should be addressed to C.-H. Chen and C.-L Wu

(E-mail: chchen@nsrrc.org.tw and clwuphys@mail.ncku.edu.tw)

**Abstract**

Hydrogen spillover and storage for single-site metal catalysts, including single-atom catalysts (SACs) and single nanocluster catalysts, have been elucidated for various supports but remain poorly understood for inert carbon supports. Here, we use synchrotron radiation-based methods to investigate the role of single-site Ti catalysts on graphene for hydrogen spillover and storage. Our *in-situ* angle-resolved photoemission spectra results demonstrate a bandgap opening and the X-ray absorption spectra reveal the formation of C−H bonds, both indicating the partial graphene hydrogenation. With increasing Ti deposition and $H_2$ exposure, the Ti atoms tend to aggregate to form nanocluster catalysts and yield 13.5% $sp^3$-hybridized carbon atoms corresponding to a hydrogen-storage capacity of 1.11 wt% (excluding the weight of the Ti nanoclusters [1]). Our results demonstrate how a simple spillover process at Ti SACs can lead to covalent hydrogen bonding on graphene, thereby providing a strategy for a rational design of carbon-supported single-site catalysts.

Chemical catalytic reactions involving hydrogen, which are ubiquitous in the chemical industry, require hydrogen molecules ($H_2$) to be dissociated into H atoms at a catalytic site. The hydrogen spillover phenomenon was first observed in 1964 in the heterogeneous catalysis system $WO_3$/Pt [2]. Many subsequent experimental results on hydrogen spillover revealed that $H_2$ could be easily dissociated on a metal catalyst (although the hydrogen-hydrogen bond in $H_2$ is strong and therefore unlikely to break under normal reaction conditions), followed by migration of the H atoms to an adjacent support surface (such as reduced metal oxides) and chemisorption there. When considering the optimization of a catalyst, common strategies to achieve the desired activity and selectivity consist in adjusting the crystallographic structure [3-4], particle size [5-10], and geometry of the catalyst [11-14]. Still, synthesizing heterogeneous catalysts with isolated active sites having efficiency at an atomic level with a clarified mechanism remains a critical challenge. Various support materials also affect catalyst optimization significantly. Some examples indicate that strong interactions with the support might enhance the catalytic performance of transition-metal catalysts via charge-transfer effects [15]. Several groups have successfully prepared transition-metal-based single-atom catalysts (SACs) to minimize the amount of metal required to catalyze the hydrogen spillover efficiently, but the catalytic activity still has much room for improvement [16,17].

Since the first report on hydrogen spillover and storage in carbon nanotubes by Dillon et al. [18], carbon supports represent an obvious alternative to metal oxides to exploit hydrogen spillover for storage, even though carbon atoms do not participate directly in the reaction as catalyst, because of their higher inertness relative to conventional oxide supports. Achieving a fundamental understanding of carbon materials as catalyst support is a challenge due to their structural and chemical complexity, which might explain the difficulties in obtaining reproducible adsorption capacities from various studies with volumetric and gravimetric measurements [19,20]. The migration of spillover-dissociated hydrogen atoms is kinetically unfavorable on graphene or graphite [21], which means that hydrogen storage on graphene supports is restricted to small distances around the catalytic sites. How many hydrogen atoms could absorb is thus an important question, which must be addressed for practical applications of H-storage. Inspired by this objective, based on our previous work showing that single Ti atoms adsorb on graphene at the energetically favorite hollow sites [22], and the structural variation upon Ti deposition on graphene and its hydrogen uptake studied by scanning tunneling microscope [23-26], we utilized Ti atomic deposition with a view to mimic SACs on graphene. Unlike previous studies, which used dissociated hydrogen sources (as shown in the Supporting Information Table S1), here we demonstrate that atomically dispersed Ti atoms on crystalline graphene can produce an efficient spillover for hydrogen, which is then chemisorbed around the anchored Ti SACs. The evaluated H storage capacity exceeds that of Ti nanocluster catalysts at a similar Ti loading.

A well-defined single-crystalline graphene support is suitable as a platform to investigate hydrogen spillover from a single-site Ti catalyst, which dissociates $H_2$ molecules into H atoms which then migrate to the support graphene (as the receptor of H atoms, schematic illustration in Fig. 1). Hydrogen atoms migrating to the support layer, assisted by the single-site Ti catalyst, can provide substantial storage of atomic hydrogen on graphene in terms of storage quantity. Graphene used in this work was epitaxially grown via thermal decomposition in a resistively heated cold-wall reactor (Aixtron HT-BM) on a hydrogen-etched 4H-SiC (0001) substrate. To achieve a sufficient monolayer uniformity, we grew pristine crystalline graphene in an Ar atmosphere at 1600 K and 780 mbar for 10 min. Single-site catalysts were formed by depositing a sub-monolayer amount of Ti atoms (0.03 to 0.5 ML), calibrated with a quartz crystal microbalance (QCM). The Ti coverage is defined with respect to hexagonal close-packed bulk Ti [23]. Angle-resolved photoemission spectra (ARPES) show a Ti adatom-induced renormalization of the band structure for low Ti coverage (~0.03 ML), as shown in Fig. 2. This was attributed to a strong Ti $3d$ and C $2p_z$ orbital hybridization of isolated Ti-atoms at the graphene hollow sites [22]. After exposing the sample to 4.5 L of $H_2$ near 300 K (1 L, exposure to $1\times10^{-6}$ Torr for 1 s), the significant ARPES spectral contrast in Fig. 2(A) shows a band gap opening of 280 meV in the graphene. This condition indicates that hydrogen atoms chemisorbed on graphene, which is evidence for the dissociation of gaseous $H_2$ molecules and hydrogen spillover by isolated Ti SACs. Moreover, without Ti, neither band structure modification nor renormalization was observed in the ARPES spectrum of graphene after $H_2$ exposure, as shown in Supporting Information Fig. S1. Thus, the Ti catalysis facilitates the dissociation of molecular $H_2$ and the hydrogenation of graphene near 300 K in vacuum.

To quantitatively understand the spillover effect on graphene, we monitored the band gaps of graphene before and after exposure to $H_2$ using energy-dispersion curves (EDC) extracted from ARPES (insets of Fig. 2 (B)). Calculations [27-29] yielded valuable information on the origin of the graphene band-structure modulations induced by C−H bonds, from fully hydrogenated insulating graphene sheets to partially hydrogenated graphene displaying semiconducting properties, with band gaps depending on the H coverage of graphene. The band gap value can thus serve to quantify the number of hydrogen atoms incorporated in partially hydrogenated graphene. Based on the DFT-based theoretical results shown in Fig. 2(B) [27-29], a quadratic relation $\Delta E_g = A_2 \eta^2 + A_1 \eta + A_0$ is obtained. $\eta$ is the absorbed hydrogen coverage on graphene (H/C value, i.e., the H storage capacity of graphene), $A_2$ = 2.49 eV, $A_1$ = 1.47 eV, and $A_0$ = 9.67 ×$10^{-2}$ eV. The band-gap opening ($\Delta E_g$) increases from about 0 to 1.5 eV as $\eta$ increases from 0 to 50 %. The H-storage capacity $\eta$ of partially hydrogenated graphene with band gap 280 meV is estimated to be about 10.6 % (0.88 wt%). As indicated in Fig. 2(B), this value is in between a lower limit of 7.3 % (0.6 wt%) obtained with the DFT+PBE approach and an upper limit of 16.2 % (1.33 wt%) with the DFT+GGA approach [27-29]. Previous theoretical and experimental work clearly showed that charge transfer between the catalyst and the support generates a noticeable impact on the electrocatalytic activity of the catalyst [30,31]. As we have concluded in our previous work [22], the anchored Ti SACs contributes a pronounced electron transfer from Ti to the graphene support (0.45 e$^-$ per Ti atom), which leads to an n-type doping of graphene. Thus, we attribute the high storage activity observed here to the charge transfer between Ti atoms and graphene, which provides a more efficient and safer method for using SACs on graphene.

To explore the nature of the formation of C−H bonds accompanying hydrogenation, we recorded X-ray absorption spectroscopy (XAS) data on Ti-decorated graphene after exposure to $H_2$, following the same sample preparation procedure as for the ARPES experiment. We recorded the XAS at an X-ray beam incident angle of 0° relative to the surface normal. The C−H* resonance region is presented in Figure 3, for Ti-doped (~0.03 ML) and $H_2$-dosed (~ 4.5 and 49.5 L) samples. We plot the spectral differences, i.e. spectra from which the spectrum of pristine graphene is subtracted. A prominent spectral feature is clearly visible centered at 287.5 eV, which is attributed to the C−H* resonance [32,33]. Its intensity increases significantly after $H_2$ dosage, while the Ti-decorated graphene (without hydrogen dosing) shows no sign of C−H bonds. This result clearly indicates the formation of C−H bond after exposing the Ti SAC-loaded graphene to hydrogen, implying the spillover of hydrogen onto the graphene support. The full-range carbon *K*-edge spectra are shown in the Supporting Information Section 4.

To seek the optimal H storage capacity, we recorded *in situ* ambient pressure X-ray photoemission spectra (APXPS) with a larger amount of Ti catalyst loading and under a larger $H_2$ exposure (1 mbar), to monitor the C−H bonding structure including the underlying substrate. This is useful to clarify their roles in the progressively hydrogenated Ti-decorated graphene under a nearly ambient gaseous pressure. Figure 4 shows the synchrotron-radiation excited C 1*s* core-level spectra recorded *in situ* near 300 K taken from pristine graphene, Ti-deposited graphene (Ti ~ 0.5 ML), and Ti-deposited graphene under $H_2$ exposure for one hour (Ti + $H_2$). By analyzing the measured C 1s core-level spectra with a Doniach-Sunjic line shape and comparing them to reference spectra of non-hydrogenated crystalline graphene grown on a SiC substrate, both measured here and from literature, the amount of hydrogen stored in our graphene samples can be precisely determined. Details of the fitting results of the C 1*s* APXPS spectra are listed in Table S3 in the Supporting Information Section 6. The C 1s core-level spectra of pristine and Ti-deposited graphene are well fitted with one graphene

sp$^3$ bonding peak (Gr at 284.4 eV), one SiC bulk component peak (B at 283.3 eV), and two interfacial components S1 and S2 related to localized reconstruction bonds, they represent a $(6\sqrt{3} \times 6\sqrt{3})R30°$ reconstructed SiC(0001) surface formed underneath the epitaxial graphene as a buffer layer [34]. A Shirley background was included in the fitting procedure. The interfacial state S1 at 284.2 eV can be identified with the covalent bonding between the buffer layer and SiC substrate, and S2 at 285.0 eV is due to the sp$^2$-bonded C atoms within the buffer layer. These two interfacial states S1 and S2 have a constant area ratio of ~ 1/2 [34,35]. They would not be affected by covering the sample with Ti catalysts and H atoms, which is consistent with the XPS results of epitaxial graphene on SiC (0001) substrate [35-37]. With Ti deposition and hydrogenation, the C 1s core level shows a progressive attenuation due to the Ti metal covering and chemisorption of atomic hydrogen. Hydrogen bonding to graphene implies that the H atom induces an out-of-plane C−H bond (Gr1, hydrogenated C atom) and can be detected in the C 1s spectrum at a higher binding energy position of 285.0 eV than sp$^2$ graphene (G) at 284.4 eV, which is consistent with the previous result of hydrogen adsorption in graphene [38]. Furthermore, another component appears in the spectrum (Gr2), which is due to non-hydrogenated C atoms adjacent to C−H bonds, as illustrated in Fig. 4. Therefore, the line shape of Gr changes gradually upon hydrogenation, and two shoulder peaks appear with respect to Gr, which are 0.6 eV higher (Gr1) and 0.2 eV lower (Gr2) in binding energy [39-42]. Thus, the Gr1 peak area is highly reliable for quantitative determination of the hydrogen content, using $\eta(\%) =$ Gr1/(Gr+Gr1+Gr2). We found here hydrogenated graphene with a H coverage of about $\eta \sim 13.5 \pm 0.5\%$ (or $1.11 \pm 0.04$ wt%), which is larger than previous reports about atomic hydrogenation of crystalline graphene ($\eta \sim 8.7$ to $9\%$) [41,43].

The hydrogen storage by spillover is a process of adsorption of atomic hydrogen, in which the numbers of catalytic sites and H adsorption sites determine the storage capacity. The density of SAC active sites, i.e., the hollow sites occupied by Ti atoms, were estimated based on the Ti coverage of 0.03 ML, which corresponds to $4.0 \times 10^{13}$ cm$^{-2}$ for the Ti SAC density. How far H atoms can migrate from the Ti catalyst is dominated by the macroscopic diffusion coefficient ($L_D$). According to a 2D-random-walk model for diffusion, $L_D$ can be expressed with a potential diagram in the classical limit. The diffusion length of H on graphene was estimated with this equation,

$$L_D = (D_0 \tau_0)^{0.5} \exp\left(\frac{E_{ad}-E_{dif}}{2k_B T}\right) \qquad (1)$$

in which $E_{ad}$ and $E_{dif}$ are the adsorption and diffusion energies of H on graphene, which are 78 KJ/mol and 91 KJ/mol, respectively [44]. $D_0$ and $\tau_0$ are the pre-exponential factors of diffusion and residence period of C–H bonds on graphene. Their values are 3276 Å$^2$/ps and 1.52 ps, respectively [45]. Near 300 K, the diffusion length $L_D$ is thus about 2.7 Å, which is only twice the C–C bond length, $a = 1.42$ Å. As the Ti catalyst atoms are anchored on the energetically favorable graphene hollow sites, the spillover H atoms diffuse isotropically and are restricted to small distances around the Ti catalytic center on the graphene surface, as depicted in Fig. 5 (A). This schematic illustration reveals that graphene hydrogenation occurs only on the first nearest-neighbor sublattices (i.e. one Ti surrounded by six H atoms). As shown in Supporting Information Section 5, the H-storage capacity $\eta$ of Ti SACs with an amount 0.03 ML is estimated to be about 6.3 %. This value is closed to the low limit value ($\eta$~7.3 %) extracted from the partially hydrogenated graphene band gap and near that of most metal catalysts on various carbon supports, especially the reported atomic metal species on carbon supports under high pressure [20,46,47].

Furthermore, when the deposition of Ti reaches 0.5 ML, H-storage capacity $\eta$ increased non-linearly to 13.5 % according to the APXPS result. This indicates that for Ti deposition above 0.1 ML, the single Ti atoms are generally susceptible to aggregate to become nanoclusters on graphene, consistent with previous STM studies [23]. For Ti nanoclusters adopting a hexagonal close-packed cluster configuration (for details, see Supporting Information Section 5) [48], the number of C−H bonding sites can be estimated to be within a concentric circular area from the nanocluster boundary to an outer diameter, which is larger by $L_D$. This is shown in Fig. 5 (A) for the case of a Ti nanocluster with Ti atom number $N$=10. With increasing Ti coverage on graphene, the number of H atoms stored around each Ti nanocluster increases sub-linearly, while the H-storage performance per Ti atom decreases sharply, as shown in Fig. 5 (B) (see also Supporting Information Table S2). Thus, due to the coalescence of Ti atoms into larger nanoclusters, the hydrogen storage capacity $\eta$ does not increase linearly with the Ti catalyst loading on graphene. Furthermore, to boost $\eta$ to 100 % on graphene, the full utilization of Ti SACs on graphene ($6.3 \times 10^{14}$ sites cm$^{-2}$ for 0.48 ML) with a Ti$_1$H$_6$ structure would need to form, as illustrated in Fig. 5 (C). Even though the H-storage capacity at ambient temperature and in vacuum reported here is still below the DOE onboard vehicular target (~7.5 wt%), simultaneously enhancing the number of Ti SACs on graphene and high-pressure H$_2$ loading is anticipated to increase the performance for practical onboard hydrogen storage for vehicular fuel cells. Also, the H-storage capacity shown here is significantly larger than all reported results of high-pressure hydrogen-storage measurements on carbon nanotubes, graphene, and Pt-decorated graphene [1,20,46,47].

In conclusion, we designed and implemented hydrogen spillover from single-atom to single-cluster catalysts by precisely controlling the Ti coverage on graphene. The atomic nature of C−H bonds on graphene and the optimal amount of H storage capacity were spectrally characterized with ARPES and XAS. Our photoemission data were quantitatively explained by a short migration distance of H atoms diffusing at room temperature on graphene. Our finding provides a general rule that will enable a rational design of carbon-supported physisorbed metal catalysts for chemisorbed H-storage via spillover.

**Supporting Information**

This material is available free of charge via the internet at http://pubs.acs.org.

Details of experimental methods, crystal growth, and additional experimental results and discussion.

**Acknowledgements**

We thank Dr. Yeu-Kuang Hwu of Academia Sinica for providing a mu-metal chamber for the ARPES measurements and NSRRC beamline staff for skillful assistance during SR experiments. This work was funded by Taiwan Ministry of Science and Technology.

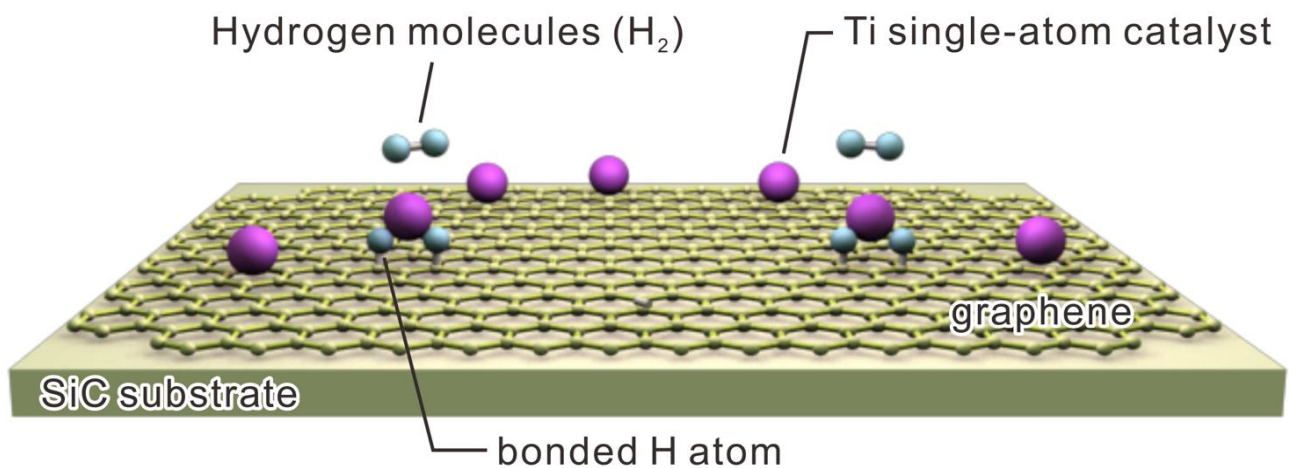

**FIG. 1 (color).** Schematic diagram of the spillover effect of Ti on graphene, which involves dissociation of molecular hydrogen near 300 K, migration of hydrogen atoms on epitaxial graphene and their chemisorption.

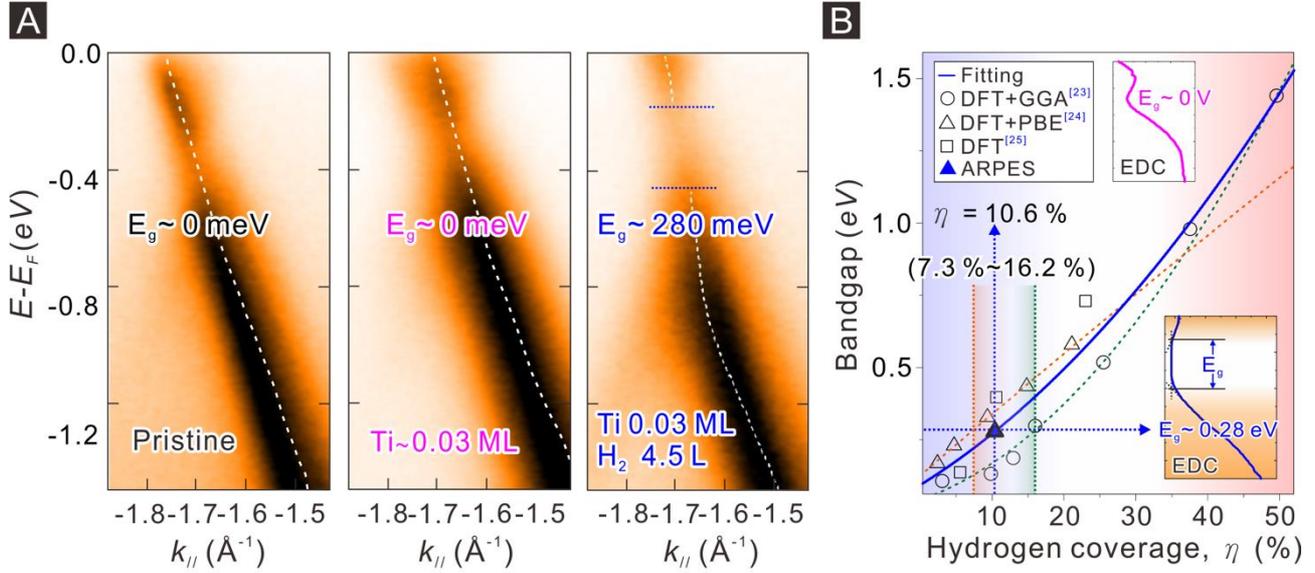

**FIG. 2 (color).** ARPES measurements performed on pristine epitaxial graphene, after Ti deposition, and after a subsequent dose of $H_2$ molecules ($H_2 = 4.5$ L) near 300 K. (A) Band structure of pristine and Ti-doped graphene for 0.03 ML. The sample was then exposed to a 4.5 L dose of $H_2$. The data were taken along direction K→Γ in the vicinity of the K-point. White dashed lines indicate the energy dispersion fitted by the momentum dispersion curves (MDC). (B) Data points from previously performed DFT calculations [27,28,29] were used to obtain the extent of graphene hydrogenation from the measured band gap induced by H-C bonds. The dependence of the bandgap of graphene on H coverage is $\Delta E_g = A_2\eta^2 + A_1\eta + A_0$, with $\eta = H/C$, $A_2 = 2.49$ eV, $A_1 = 1.47$ eV, and $A_0 = 9.67 \times 10^{-2}$ eV. The hydrogen coverage η corresponding to graphene band gap $E_g=0.28$ eV is obtained via two different approaches, with a lower limit of 7.3 % and an upper limit of 16.2 % using a fit of DFT+PBE and DFT+GGA data, respectively. Insets show angle-integrated curves of energy distribution (EDC) in the vicinity of the K point taken before (red) and after (blue) 4.5 L $H_2$ dosage for graphene with 0.03 ML Ti.

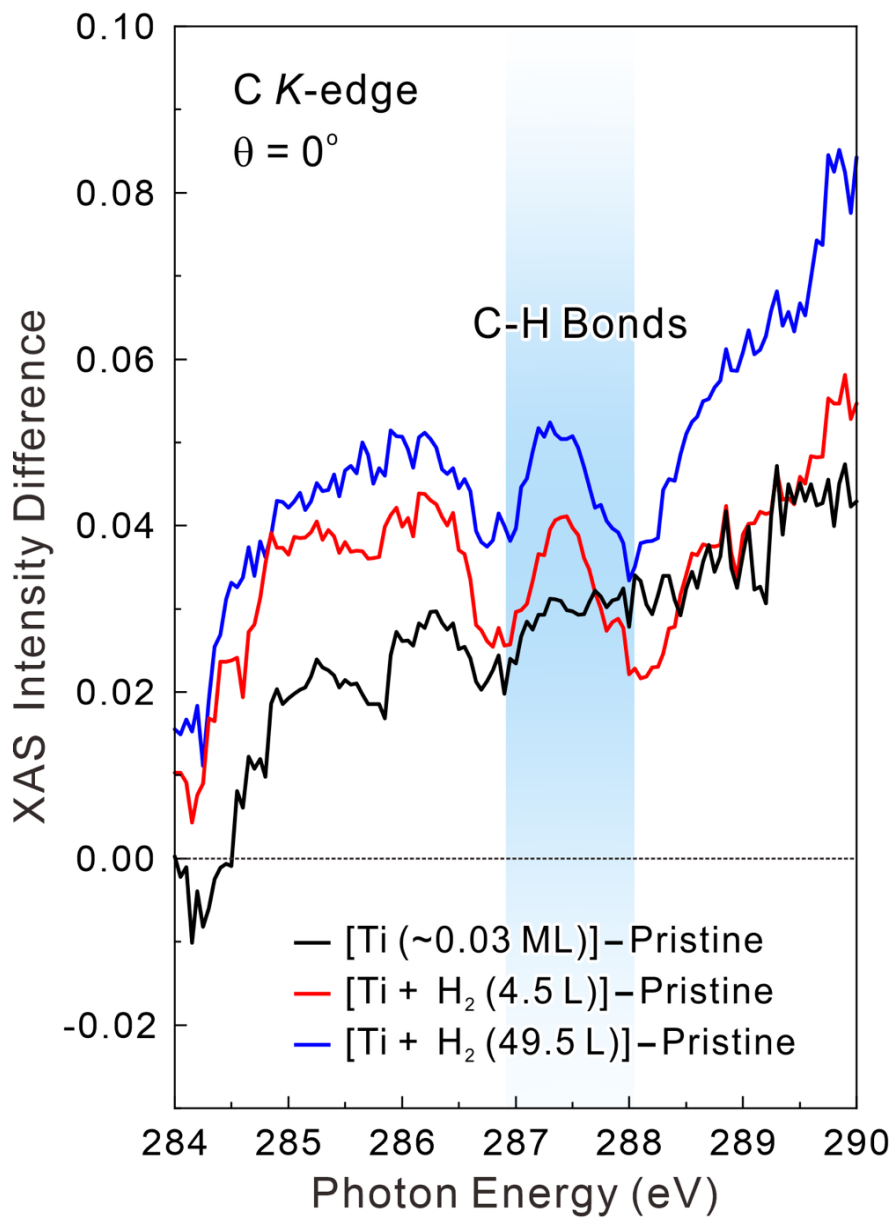

**FIG. 3 (color).** C−H* resonance region of the Carbon *K*-edge X-ray absorption spectra for Ti-doped (black curve), for 4.5 L $H_2$-dosed (red curve), and for 49.5 L $H_2$-dosed (blue curve) samples. We show difference spectra, i.e., spectra minus the spectrum of pristine graphene. The shaded area indicates the position of the C−H* resonance. See Supporting Information Section 3 for full spectra.

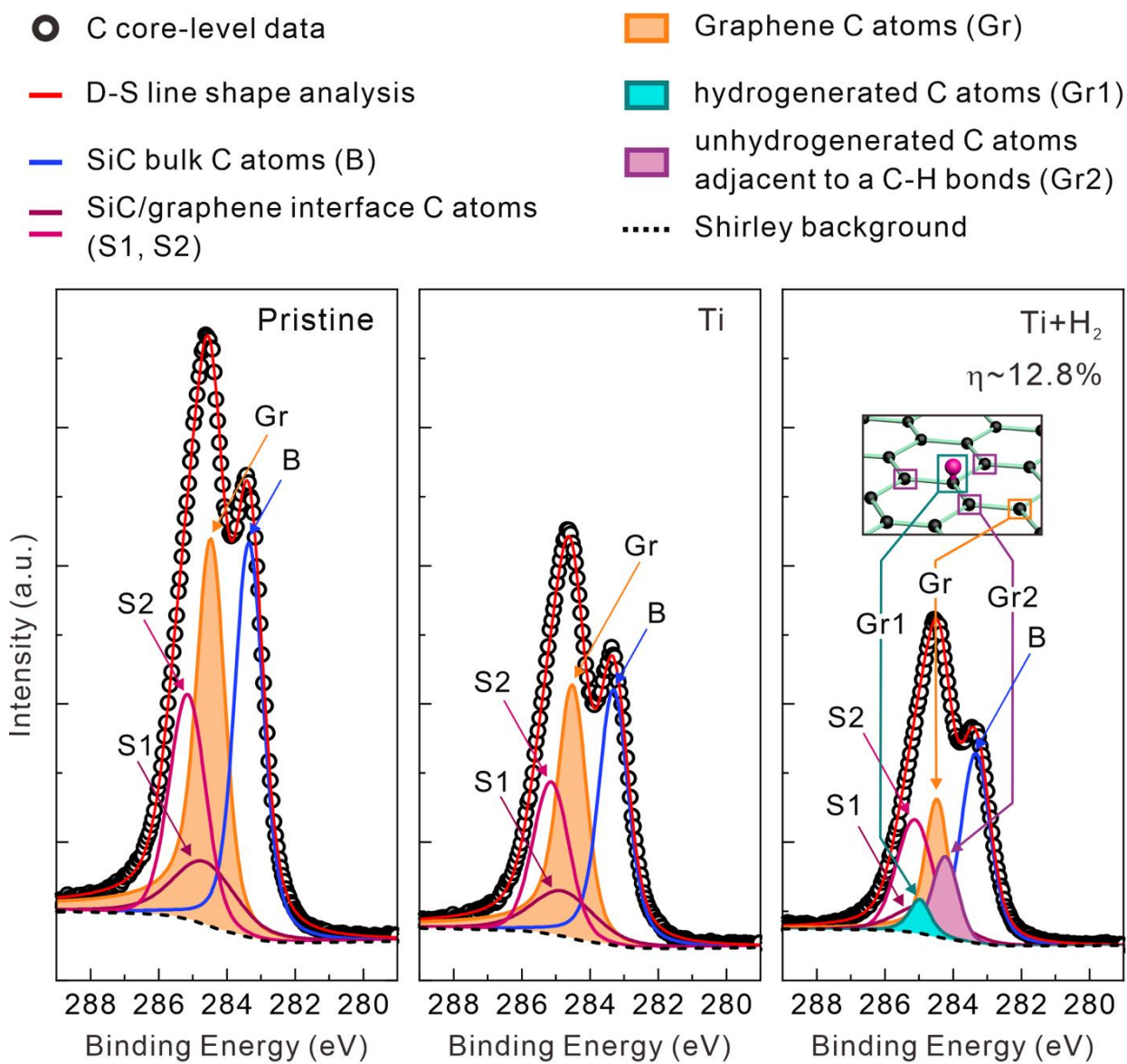

**FIG. 4 (color).** From left to right: APXPS C 1s spectra of pristine, Ti-deposited (0.5 ML), and Ti + $H_2$ exposed graphene ($H_2$ 1 mbar for 1 hour). Spectra were deconvoluted using Gaussian-broadened Doniach-Sunjic (DS) functions including Shirley background. Gr, B, and S1, S2 indicate C 1s chemical environments for graphene, SiC bulk, and interfacial reconstruction (at the buffer layer interface), respectively. Component Gr1 represents hydrogenated C atoms (C–H bond) of graphene. Gr2 represents not-hydrogenated C atoms next to C–H bonds, as illustrated in the inset. From the integrated area of the relative constituents of the graphene C 1s C–H components (Gr, Gr1, Gr2), we estimate $\eta$~13.5 % hydrogen coverage.

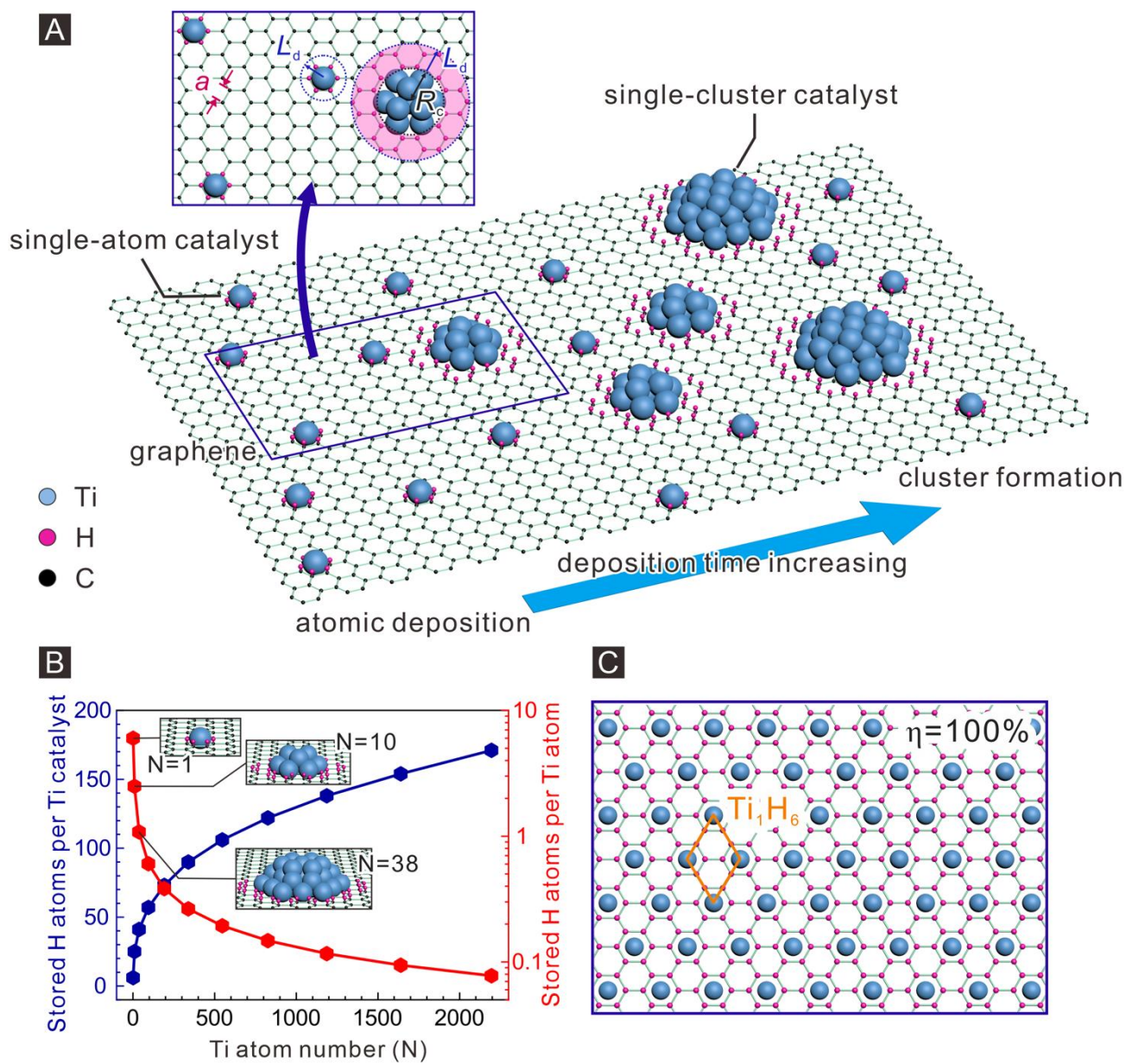

**FIG. 5 (color).** (A) Schematic diagram of single-atom to single-cluster catalyst formation. Low to high Ti deposition was explored to obtain well-isolated atoms or nanocluster catalysts on graphene. From the spillover of H atoms for an isolated Ti atom and Ti nanocluster catalyst, the 2D-diffusion length ($L_D$) and nanocluster size (diameter $R_C$) can be used to estimate the H-storage capacity for small and large Ti coverages. (B) Calculated H storage of Ti SACs and Ti nanoclusters with truncated bipyramid shape versus number $N$ of Ti atoms. (C) Schematic illustration of full H storage ($\eta=100$ %) on graphene with the help of Ti SACs through the formation of a $Ti_1H_6$ structure.

ToC Figure:

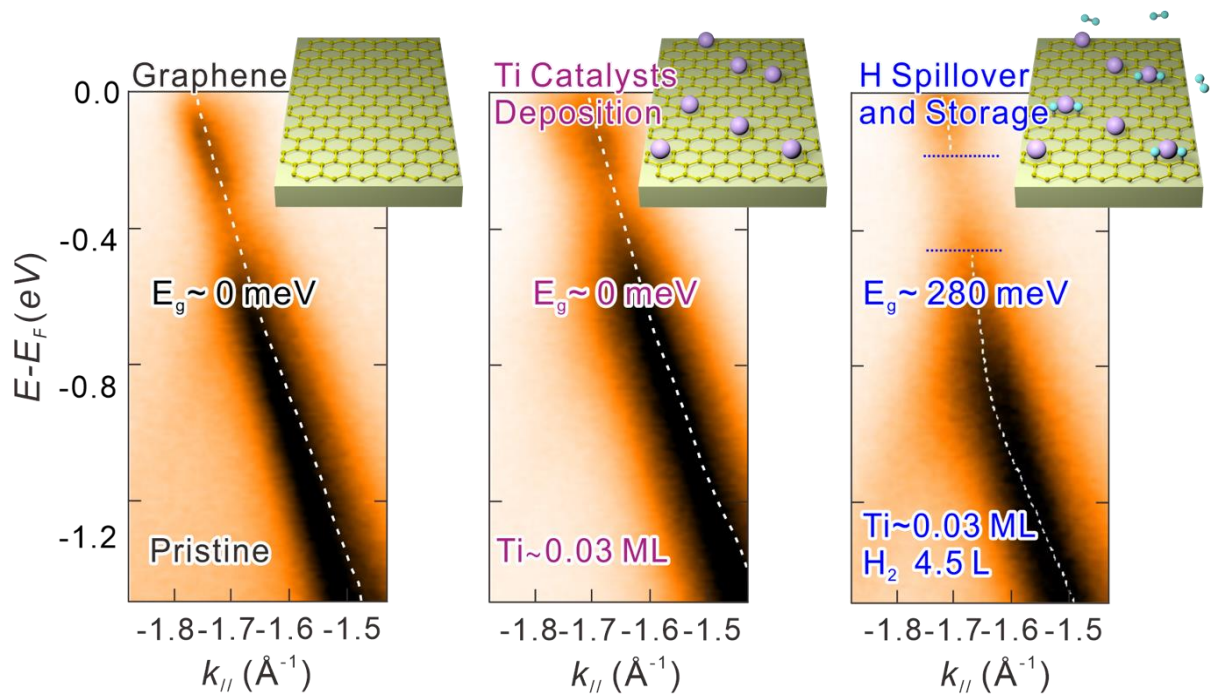

*Supporting Information for*

# Hydrogen spillover and storage on graphene with single-site Ti catalysts

Jhih-Wei Chen[1,2], Shang-Hsien Hsieh[2], Sheng-Shong Wong[1], Ya-Chi Chiu[1], Hung-Wei Shiu[2], Chia-Hsin Wang[2], Yaw-Wen Yang[2], Yao-Jane Hsu[2], Domenica Convertino[3], Camilla Coletti[3], Stefan Heun[4], Chia-Hao Chen[2*], and Chung-Lin Wu[1,2*]

[1]Department of Physics, National Cheng Kung University, Tainan 70101, Taiwan

[2]National Synchrotron Radiation Research Center, Hsinchu 30076, Taiwan

[3]Center for Nanotechnology Innovation @ NEST, Istituto Italiano di Tecnologia, Pisa 56127, Italy

[4]NEST, Istituto Nanoscienze-CNR and Scuola Normale Superiore, 56127 Pisa, Italy

## 1. Experimental methods

**Crystal growth**

Graphene was grown via thermal decomposition in a resistively heated cold-wall reactor (Aixtron HT-BM) on a hydrogen-etched 4H-SiC (0001) substrate. The growth was performed under Ar atmosphere, at 1600 K and 780 mbar for 10 min. The quality and homogeneity of the monolayer graphene was assessed with a combined analysis of Raman spectra and atomic-force microscopy. Before ARPES, XAS, and APXPS measurements, the graphene samples were annealed at 900 K for more than 7 h under UHV conditions.

**Details of measurements**

ARPES was used to probe the electronic structure of pristine graphene and of Ti-decorated graphene, with and without the dosage of $H_2$ molecules. APXPS measurements were performed to determine the percentage of hydrogenation of Ti-decorated graphene as $H_2$ molecules adsorbed. After dosing the sample under continuous flow of $H_2$ for various time periods, the ARPES and APXPS data were measured after the chamber was pumped down to the base pressure. Analyzers (SPECS Phoibos 150 and SPECS Phoibos 150 NAP) were utilized to measure the ARPES and APXPS spectra, respectively. Measurements were performed at Taiwan Light Source (TLS) beamlines 08A1 and 24A1 of the National Synchrotron Radiation Research Center (NSRRC) in Hsinchu, Taiwan. All measurements were performed near 300 K. Energy and momentum resolution are estimated to be better than 150 meV and $\pm\,0.005$ Å$^{-1}$ for ARPES. Energy resolution is estimated to be better than 280 meV for APXPS. The XPS spectra were deconvoluted with the software UNIFIT 2013. The peak area uncertainties were carefully analyzed according to R. Hesse et. al. [1].

The XAS spectra were recorded at TLS beamline 09A2. The incident angles of the linearly polarized undulator radiation were 0° and 45° with respect to the graphene surface normal. We recorded the sample current to obtain the XAS spectra in the total-electron-yield mode. All XAS spectra were recorded in on-the-fly scan mode to prevent radiation damage. The photon energy was calibrated against the 1s–π* resonance of HOPG at 285.5 eV.

A QCM was used to determine the Ti coverage, as a QCM has a narrow resonance that leads to highly stable oscillations and great accuracy, even in the case of minute coverages. To duplicate the deposition conditions, we placed the QCM in front of the Ti evaporator at the same distance, deposition angle and base pressure as the sample.

**References**

1. Hesse, R., Chassé, T., Streubel, P., and Szargan, A. R. Error estimation in peak-shape analysis of XPS core-level spectra using UNIFIT 2003: how significant are the results of peak fits? Surf. Interface Anal. **36,** 1373 (2004)

## 2. Band diagrams measured by ARPES on bilayer graphene (BLG) and H$_2$-exposed BLG (without Ti).

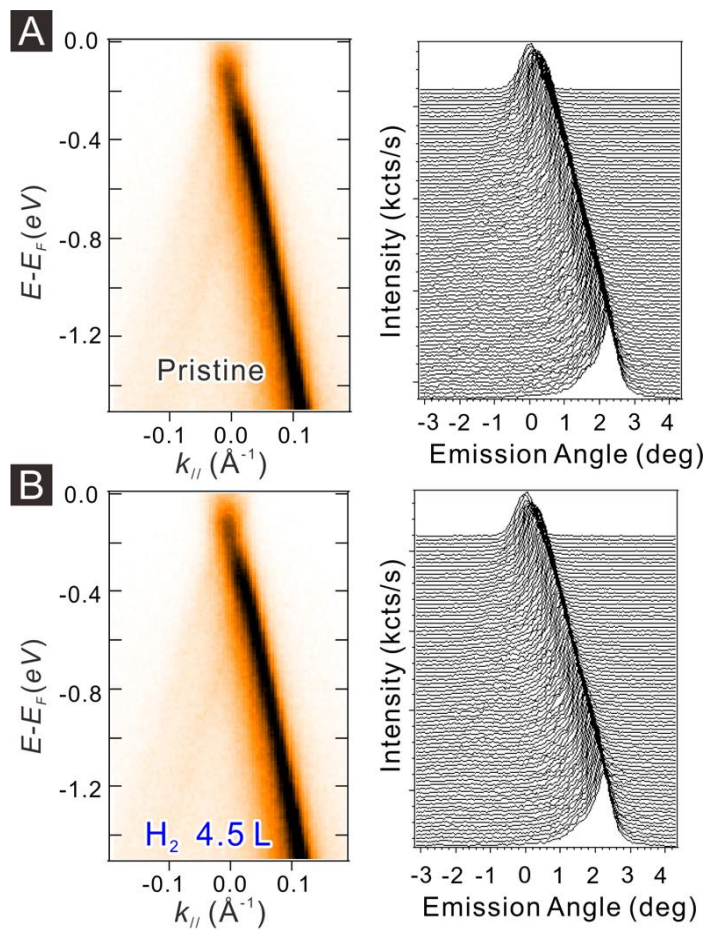

*Figure S1*: *ARPES (left) and MDC (right) measured on (A) pristine bilayer graphene and (B) after exposure to 4.5 L of molecular hydrogen. No Ti was deposited prior to hydrogenation. The results show that the addition of molecular hydrogen near 300 K does not alter the band structure of bilayer graphene (without Ti).*

# 3. Summary of experimental hydrogen storage capacity in graphene

**Table S1**

**Table 1**

**Summary of experimental hydrogen storage capacity in graphene**

| H source | support | catalyst | T (k) | Pressure | Capacity (% and wt %) | | Ref. |
|---|---|---|---|---|---|---|---|
| $H_2$ plasma | Exfoliated graphene | No | RT | 1 torr | 16.67 % | | 1 |
| $H_2$ plasma | CVD graphene | No | RT | 0.015 torr | 33 % | | 2 |
| $H_2$ plasma | CVD graphene | No | RT | 0.01 torr | 25 % | | 3 |
| cracked $H_2$ | CVD graphene on Ni (111) | No | RT | $1.5 \times 10^{-6}$ torr | 25 % | | 4 |
| cracked $H_2$ | CVD graphene on Ni (111) | No | RT | $1.5 \times 10^{-6}$ torr | 25.6 % | | 5 |
| cracked $H_2$ | CVD graphene on Ni (111) | No | RT | $1.5 \times 10^{-9}$ torr | 25 % | | 6 |
| $H_2$ | Exfoliated graphene | No | RT | 6 MPa | | 0.2 wt % | 7 |
| $H_2$ | Graphene powder | No | RT | 10 bar | | 0.72 wt % | 8 |
| $H_2$ | Nitrogen-doped graphene | Pd | RT | 4 MPa | | 4.4 wt % | 9 |
| $H_2$ | Nitrogen-doped graphene | Pd NPs | RT | 2 MPa | | 1.9 wt % | 10 |
| $H_2$ | Nitrogen-doped graphene | Pd NPs | RT | 4 MPa | | 4.3 wt % | 11 |
| **$H_2$** | **Crystalline graphene on SiC (0001)** | **Ti** | **RT** | **1 mbar (1 hr)** | **13.5 %** | **1.11 wt %** | **This work** |

## 4. Carbon *K*-edge XAS spectra

Polarization-dependent carbon *K*-edge XAS spectra taken with different soft X-ray incident angles relative to the surface normal are shown in Figure S2. For well-ordered systems, the polarization-dependent XAS can serve to probe the orientation of molecular orbitals [1,2]. The resonance intensity associated with a particular molecular orbital is largest when the electric field of the incoming light is parallel to the direction of the orbital, and is minimized when the field is perpendicular to the orbital. Namely, normal incident soft X-rays (i.e., 0° with respect to the surface normal) are only sensitive to in-plane orbitals, whereas 45° incident soft X-ray are sensitive to both in-plane and out-of-plane orbitals. Since the π orbital of the graphene sheet is perpendicular to the surface, only the spectra with 45° incidence can reveal the π* resonance.

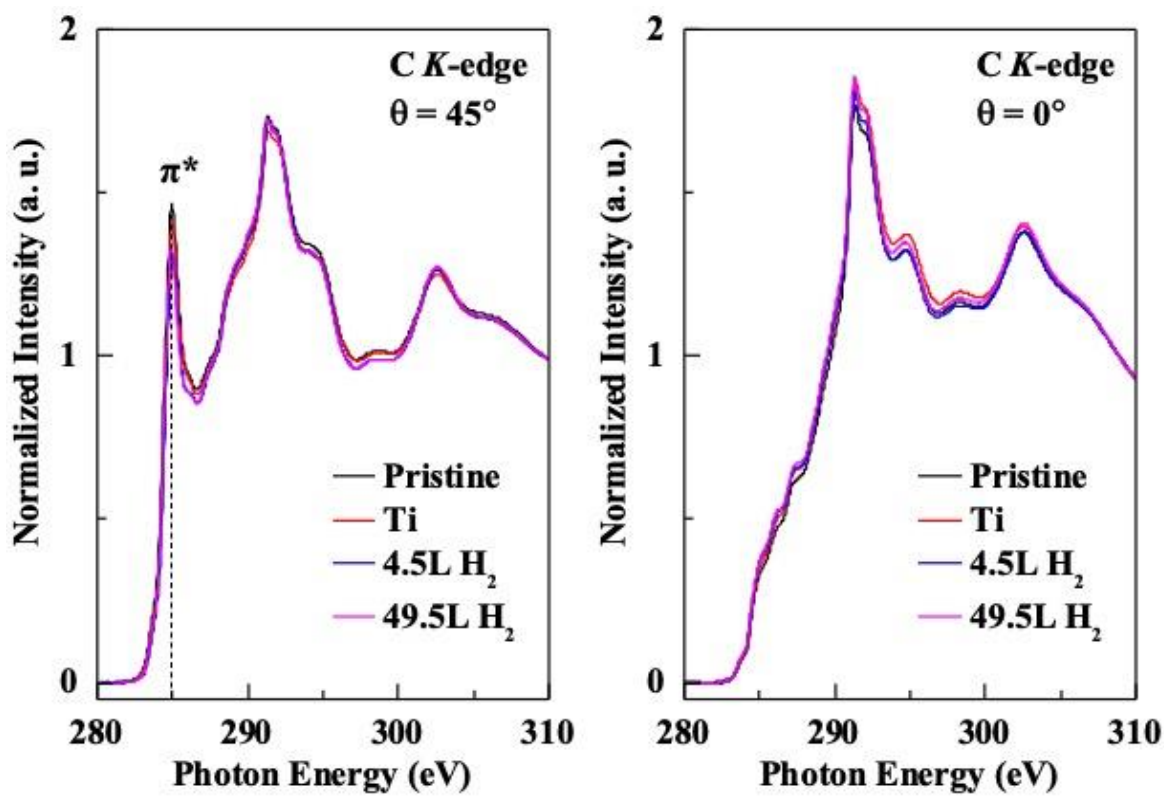

*Figure S2*: *The full range XAS spectra of pristine graphene, Ti doped, low $H_2$ dosage (4.5L), and high $H_2$ dosage (49.5L) samples. The indicated angles are the angles of the incident soft X-ray beam relative to the surface normal.*

Further, by taking the spectral difference, i.e., the 45° spectra minus the 0° spectra, we will effectively select the out-of-plane orbital alone. Figure S3 shows the difference spectra (45°− 0°) of the π* resonance region, for pristine, Ti-doped (~0.03 ML), and $H_2$-dosed (~ 4.5 to 49.5 L) graphene. A pronounced asymmetric π* resonance as a spectral feature is clearly visible at about 285.5 eV. Its intensity decreases significantly with increasing $H_2$ dosage. This result indicates the formation of C–H bonds that suppress the number of π orbitals.

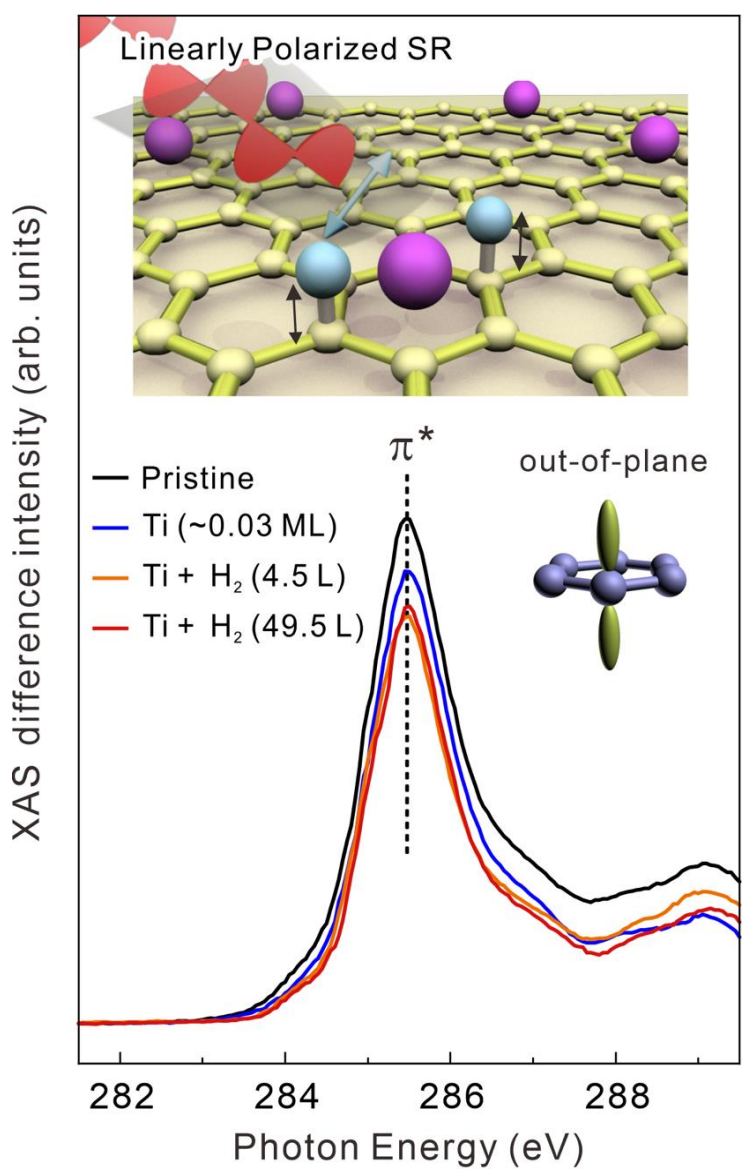

*Figure S3*: *Carbon K-edge X-ray absorption spectra measured with linearly polarized soft X-rays on pristine graphene, after Ti deposition (Ti ~ 0.03 ML), and after exposure to molecular hydrogen*

*$H_2$ (4.5 and 49.5 L). We recorded the XAS spectra at 0° and 45° relative to the surface normal (see Fig. S2). The difference spectra (45°-0°) are presented here. The schematic diagram shows the resonance characteristics corresponding to a C-C $\pi$\* orbital (~ 285.5 eV) on graphene.*

## 5. H storage capacity of Ti SACs and Ti nanoclusters of different size on graphene

a. For a single-site Ti atom catalyst,

0.03 ML = $4.0 \times 10^{13}$ cm$^{-2}$

$$\eta_H = \frac{H}{C} = \frac{6 \times (4.0 \times 10^{13})}{3.8 \times 10^{15}} \times 100\% \approx 6.3\%$$

b. For a single-site Ti nanocluster catalyst,

**Table S2: Details of calculation of hydrogen storage capacity**

| Number N of Ti atom in nanoclusters of different size [1] | Nanocrystal Radius R$_c$ (nm) [2] | Diffusion area (nm$^2$) | H storage amount | H storage per Ti atom |
|---|---|---|---|---|
| 10 | 0.251 | 0.229 | 25 | 2.5 |
| 38 | 0.502 | 0.655 | 41 | 1.079 |
| 95 | 0.753 | 1.081 | 57 | 0.6 |
| 192 | 1.005 | 1.507 | 73 | 0.38 |
| 339 | 1.256 | 1.933 | 90 | 0.265 |
| 547 | 1.507 | 2.359 | 106 | 0.194 |
| 826 | 1.758 | 2.785 | 122 | 0.148 |
| 1187 | 2.009 | 3.211 | 138 | 0.116 |
| 1640 | 2.26 | 3.638 | 154 | 0.094 |
| 2196 | 2.511 | 4.064 | 171 | 0.078 |

Hydrogen storage amount is calculated as H storage amount = Diffusion area × C–H site density

H storage per Ti atom is calculated as H storage per Ti atom = $\frac{\text{Diffusion area} \times \text{C–H site density}}{N}$

# 6. Formation of C–H bonds on graphene/SiC and its identification using APXPS

From the integrated area of the Gr1 component, we can determine the hydrogen coverage (in %) of crystalline graphene with Ti nanocluster catalysts.

**Table S3** Summary of fitting peak position, relative intensity, and relative area of every peak of the C 1s APXPS spectra (pristine, Ti, and Ti+$H_2$) after baseline subtraction.

|  | peak | Position (eV) | Intensity (a.u.) | Area |
|---|---|---|---|---|
| **Pristine** | Gr | 284.4 | 56141.7 | 80760 |
|  | B | 283.4 | 57071.4 | 70301 |
|  | S1 | 284.7 | 9098.9 | 24696 |
|  | S2 | 285.2 | 32528.5 | 46114 |
| **Ti** | Gr | 284.5 | 36700.0 | 52768 |
|  | B | 283.3 | 37000.6 | 47492 |
|  | S1 | 284.8 | 5982.1 | 16239 |
|  | S2 | 285.1 | 22000.2 | 31193 |
| **Ti+$H_2$** | Gr | 284.4 | 20000.0 | 22187 |
|  | B | 283.3 | 27600.1 | 34834 |
|  | S1 | 284.7 | 4500.1 | 11563 |
|  | S2 | 285.1 | 16499.9 | 23302 |
|  | Gr1 | 285.0 | 5100 | 5378 |
|  | Gr2 | 284.1 | 15000 | 12189 |